# Sulfuric acid as a cryofluid and oxygen isotope reservoir of planetesimals

Akihiko Hashimoto[1] & Yuki Nakano[2]*

[1]Suns & Fishermen Society, Sapporo 060-0001, Japan.
[2]Institute of Low Temperature Science, Hokkaido University, Sapporo 060-0819, Japan.
*Corresponding author, E-mail: ynakano@lowtem.hokudai.ac.jp

**The Sun exhibits a depletion in $^{17,18}$O relative to $^{16}$O by 6 % compared to the Earth and Moon[1]. The origin of such a non-mass-dependent isotope fractionation has been extensively debated since the three-isotope-analysis[2] became available in 1970's. Self-shielding[3,4] of CO molecules against UV photons in the solar system's parent molecular cloud has been suggested as a source of the non-mass-dependent effect, in which a $^{17,18}$O-enriched oxygen was trapped by ice and selectively incorporated as water into planet-forming materials[5]. The truth is that the Earth-Moon and other planetary objects deviate positively from the Sun by ~6 % in their isotopic compositions. A stunning exception is the magnetite/sulfide symplectite found in Acfer 094 meteorite, which shows 24 % enrichment in $^{17,18}$O relative to the Sun[6]. Water does not explain the enrichment this high. Here we show that the SO and $SO_2$ molecules in the molecular cloud, ~106 % enriched in $^{17,18}$O relative to the Sun, evolved through the protoplanetary disk and planetesimal stages to become a sulfuric acid, 24 % enriched in $^{17,18}$O. The sulfuric acid provided a cryofluid environment in the planetesimal and by itself reacted with ferric iron to form an amorphous ferric-hydroxysulfate-hydrate, which eventually decomposed into the symplectite by shock. We indicate that the Acfer-094 symplectite and its progenitor, sulfuric acid, is strongly coupled with the material evolution in the solar system since the days of our molecular cloud.**

The GENESIS mission has successfully brought a solar wind for us to determine the Sun's oxygen isotopic composition: $\delta^{18}O \sim \delta^{17}O \sim -60$ ‰[1]. The value must be identical to the average oxygen isotopic composition of the parent molecular cloud from which our solar system evolved, because the Sun represents more than 99% of its total mass. Then, why do the Earth and Moon system, our world, as well as all of the fragments (meteorites) from planetary and pre-planetary bodies (planetesimals) in the solar system exhibit oxygen isotopic compositions so different from that of the Sun (Fig. 1)? Even before the solar wind mission, a puzzling mineral assemblage, a symplectic intergrowth of magnetite/sulfide named "cosmic symplectite"[7] was known to exist in a unique class of carbonaceous chondrite, Acfer 094. Magnetite in this symplectic assemblage exhibits $\delta^{18}O \sim \delta^{17}O \sim +180$ ‰[6] which occupies its position on the opposite side of the Sun (Fig. 1).

Oxygen isotopic composition of a substance is expressed as the number ratios of isotopes, $^{18}$O and $^{17}$O to $^{16}$O, defined by $\delta^{i}O = ((^{i}O/^{16}O)_{sp} /(^{i}O/^{16}O)_{std} - 1)) \times 1000$ ‰, where i= 18 or 17, sp: substance, std: Standard Mean Ocean Water (SMOW), and ‰: per mil. The differences in oxygen isotopic composition among various solar system materials and the Sun as shown in Fig. 1 cannot be explained by isotope fractionations which accompany ordinary chemical reactions or kinetic processes, that would shift isotopic compositions only along a slope one half line and is called mass-dependent isotope fractionation. To explain a genetic relationship between two substances whose isotopic compositions do not fall on the same slope 1/2 line, some isotope fractionation process which behaves non-mass-dependently is required. Otherwise, they must be regarded isotopically incompatible, viz. unable to evolve from the other.

To resolve the problem of the oxygen isotopic incompatibilities, studies on discrete reaction kinetics that induce non-mass-dependent isotope fractionations[8,9] have been applied to various gas and solid reactions pertinent to the protoplanetary environment, and are still valid working hypotheses. On the other hand, the self-shielding of CO isotopologues against photodissociation has been known to be active in extant molecular clouds[3,4], which is the most promising mechanism to create a large oxygen isotope heterogeneity because of a high abundance of CO molecules.

According to the CO self-shielding hypothesis, an isotopically single component, viz. a molecular cloud having a uniform oxygen isotopic composition can evolve to three isotopically distinct components, namely the $^{16}$O-rich CO gas molecules, the $^{17,18}$O-rich O atoms, and the rest either in the gas or solid state (mostly $H_2O$ ice and silicates) that keeps the original oxygen isotopic composition of the cloud (Fig. 2a). The O atoms of photodissociation origin would combine with H atoms on the surface of $H_2O$ ice at low temperatures (<40K) and form adsorbed $H_2O$ molecules. The latter would mix with the basal $H_2O$ ice having the original isotopic composition, resulting in the $H_2O$ ice with an intermediate oxygen isotopic composition ($\delta^{18}O \sim \delta^{17}O \sim +70$ ‰[5] or +50 ‰[10]).

Magnetite typically occurs as a hydrothermal precipitate in terrestrial and extraterrestrial environments, and for this reason the anomalous isotopic composition of the symplectic assemblage in Acfer 094 has been attributed to some unspecified "primordial water" in the protoplanetary disk[6,7]. Its value, $\delta^{18}O \sim \delta^{17}O \sim +180$ ‰, however, is much larger than the theoretical upper limit for the $H_2O$ ice described above. The initial water before its reacting with silicate minerals in many carbonaceous meteorites is estimated to have had $\Delta^{17}O < 10$ ‰[11,12], while that of the magnetite in the Acfer-094 symplectite is ~90 ‰. (The notation $\Delta^{17}O$, defined by $\Delta^{17}O = \delta^{17}O - 0.52 \times \delta^{18}O$, is often used to remove a possible mass-dependent effect and signify isotopic incompatibility among substances.)

Here we call the symplectic assemblage in Acfer 094 "Acfer-094 symplectite" because symplectites are not uncommon among extraterrestrial materials. Since $H_2O$ is the most abundant O-bearing molecule in protoplanetary disks and molecular clouds, and is reactive with most substances including silicates, it cannot be a major source of the anomalous oxygen isotopes in the Acfer-094 symplectite.

We suggest $H_2SO_4$ (sulfuric acid) for the origin of oxygen in the Acfer-094 symplectite, based on four lines of evidence. A petrographic relationship of the symplectite to the Acfer 094 meteorite matrix is interrogated for an additional clue to its origin. We argue that its anomalously $^{17,18}$O-rich composition originated from the SO or $SO_2$ molecules that incorporated heavy O atoms of photodissociation origin, $\delta^{18}O \sim \delta^{17}O \sim 1000$ ‰, in the parent molecular cloud of the solar system. They evolved to $H_2SO_4$ by reaction with OH or $H_2O_2$ in the solar protoplanetary disk. In the Acfer 094's parent planetesimal, a highly acidic and oxidizing fluid represented by the $H_2SO_4$-$H_2O_2$-$H_2O$ system provided a cryo-alteration environment for metal, sulfides, and silicates to partially dissolve. A colloidal ferric hydroxyl sulfate, $Fe_4SO_4(OH)_{10} \cdot nH_2O$ (n=1-3) precipitated from the cryofluid, which later decomposed into the symplectic assemblage during a shock event.

**$H_2SO_4$ as a carrier of heavy O-isotopes**

*The first line of evidence*

The Acfer-094 symplectite is a vermicular symplectite of iron-nickel sulfide (pyrrhotite or pentlandite) and magnetite with an approximate (Fe,Ni)S : $Fe_3O_4$ = 1 : 1 molar ratio, viz., (Fe+Ni) : S : O = 4 : 1 : 4[6,7,13]. Both magnetite and sulfide grains are elongated in shape and (10-20) x (100-300) nm in size[7,13]. They repeat similar textural patterns and crystallographic orientations[7] in larger scales, indicating a typical symplectite. Origin of many symplectites in nature is a coherent decomposition of a stoichiometric single phase into a combination of more stable phases upon a sudden change in temperature and/or pressure. We looked for such an oxide phase that had a chemical stoichiometry of Fe : S = 4 : 1. To the best of our knowledge, candidates are (a) amorphous ferric hydroxyl sulfates[14,15], $Fe_4SO_4(OH)_{10} \cdot nH_2O$ (n=1-3), (b) volaschioite[16], $Fe_4(SO_4)O_2(OH)_6 \cdot 2H_2O$, or (c) an $SO_4^{2-}$-rich extreme of schwertmannite[17], $Fe_4O_4(OH)_2(SO_4)(H_2O)_n$.

The occurrence of a natural mineral ("glockerite") having the chemical formula, $Fe_4SO_4(OH)_{10}$, is discredited now[18]. Many synthetic experiments, however, have readily produced amorphous basic ferric sulfates, $2Fe_2O_3 \cdot SO_3 \cdot nH_2O$ (n≥6)[14] or $Fe_4SO_4(OH)_{10} \cdot nH_2O$ (n=0-3)[15,19-21], as precipitates from various ferric sulfate solutions. The presence of OH ligands in the amorphous basic ferric sulfate was meticulously illustrated by the titration experiment of ferric ammonium sulfate with a free sulfuric acid using a known concentration of NaOH solution as a titrant[19]. After neutralization of the free sulfuric acid by NaOH, a precipitation of a basic ferric sulfate started with the $NaOH_{ex}/Fe^{3+}$ concentration ratio X = 0.4 (pH ~ 2.5), where $NaOH_{ex}$ represents excess NaOH after the neutralization. The precipitation continued until X ~ 2.2 with a nearly constant pH ~ 3, and completed with X = 2.49 (pH ~ 4.5). The value X = 2.49 is interpreted as the $OH/Fe^{3+}$ ratio of the precipitate, indicating that it has a chemical formula with Fe : OH = 4 : 10, thus $Fe_4SO_4(OH)_{10} \cdot nH_2O$.

In another titration experiment[14] of ferric sulfate with NaOH, natrojarosite $NaFe_3(SO_4)_2(OH)_6$ precipitated with X < 1, due probably to the absence of ammonium buffer unlike in the previous one[19]. Natrojarosite was the only phase with X < 1. With 1 < X < 2, amorphous ferric sulfate precipitated. With X > 2, the amorphous ferric sulfate was gradually replaced by goethite, α-FeO(OH). Yet, another experiment[22] showed clearly the order of precipitation in an alkalis-adding ferric sulfate solution that jarosite (pH<3), schwertmannite (3<pH<6), and ferrihydrite or goethite (pH>6) precipitated as pH is systematically increased.

Here, we postulate the amorphous ferric hydroxysulfate hydrate with the chemical formula $Fe_4SO_4(OH)_{10} \cdot nH_2O$ (n=1-3) for a possible precursor of the Acfer-094 symplectite, and temporarily call it "g-phase" because it is a colloidal precipitate[20,21] or gel-like and also "g" is reminiscent of glockerite. We leave crystalline volaschioite and schwertmannite as a possible precursor of the Acfer-094 symplectite.

The g-phase most likely formed by precipitation from a ferric-iron-rich, sulfate solution that briefly existed under the surface of Acfer 094's primary parental body, one of hypothetical planetesimals that existed in numerous numbers before they were assembled into less but large planets. In order to produce ferric-iron-rich sulfate solutions, Fe-Ni metal and/or sulfides (Fe>Ni) were available for raw materials as they do exist in the matrix of Acfer 094, in addition to the need for sulfuric acid. A strong oxidizer was also needed to oxidize ferrous iron ($Fe^{2+}$) into ferric iron ($Fe^{3+}$) in the solution. The coexistence of hydrogen peroxide ($H_2O_2$) in the sulfuric acid solution is predicted in the third line of evidence to follow.

The g-phase must have decomposed in situ and in a fraction of second to become a nano-textured symplectite in order to arrest diffusion of elements. Meteorites unequivocally have records of shock, of which origin, extent, and frequency are diverse[23]. Acfer 094 meteorite has a shock record[24] assigned as S1, that implies shock pressure[25] of < 4-5 GPa. Regardless of the weakest shock (or unshocked state) assigned for the bulk Acfer 094 meteorite, its matrix contains sulfides (mostly pyrrhotite) with severe planar fractures[6,13]. Curiously these fractured sulfides are accompanied by the symplectite in most cases, while majority of sulfide grains unaccompanied by the symplectite are fracture-free[13]. Shock experiments performed on pyrrhotite (mixture of hexagonal and monoclinic crystals)[26] indicated that planar deformation and fractures dominated up to 8 GPa, whereas at higher shock pressures large crystals developed along with amorphous or molten sulfides. While the presence of severely fractured sulfide grains in Acfer 094 suggests a shock pressure up to ≤ 8 GPa, Acfer 094 is virtually unshocked. This inconsistency is solved if shock impedance mismatching is considered.

Shock impedance, defined as a product of pre-shock density multiplied by shock wave velocity, is large for mineral grains with high density and high hardness[27]. As a result of pairing mineral grains, one with a large impedance and the other with a low impedance compared to that of their surrounding material (here the Acfer 094 matrix), their interface produces reverberated waves upon shock, inducing a shock pressure on the large impedance side higher than the ambient shock pressure, while the low impedance side is subjected to a lower shock pressure[28]. The larger the impedance contrast (viz., mismatch), the larger the shock pressure on the large impedance side. Pyrrhotite has a density, ~4.62 $g/cm^3$, and large bulk and shear moduli[29] (indicators of hardness, directly related to elastic wave velocities) of 143 and 59 GPa, respectively. The g-phase is expected to be very soft and much less dense since it is a colloidal precipitate. The expected outcome would be a high shock pressure and a high postshock temperature in the pyrrhotite grains for an insignificant shock loading on the whole rock, S1 for Acfer 094. The g-phase, on the other hand, would be much less shocked, but exposed to a high temperature because it was attached to the pyrrhotite.

We postulate that a sudden temperature rise exerted on the g-phase prompted it to dehydrate. Weak bonds (i.e., -$OH_2$ and -OH) in the g-phase must have broken in such a fashion: $Fe_4(SO_4)(OH)_{10} \cdot nH_2O \rightarrow (Fe_4(SO_4))^* + 10OH + nH_2O$, where asterisk (*) denotes a transition state. Then, a symplectic decomposition must have followed: $(Fe_4(SO_4))^* \rightarrow FeS + Fe_3O_4$. This would explain the 1 : 1 molar ratio of sulfide and magnetite observed in the Acfer-094 symplectite.

Recently, a single grain in Acfer 094 among five of those studied three-dimensionally has revealed a layered structure[30] with a central single crystal of $Na_2SO_4$, surrounded sequentially by a coarse- to fine-grained symplectic assemblage of magnetite and iron sulfide and an outermost layer of iron oxide. The $Na_2SO_4$ exhibits oxygen isotopic composition[30] similar to that of the coexisting symplectic magnetite and to that of the Acfer-094 symplectite originally characterized[6].

The mineralogical sequence, from core to rim, in this newly found structure mirrors the precipitation order of minerals from a Na-containing ferric sulfate solution with increasing pH described above, if $Na_2SO_4$, the symplectite, and iron oxide are replaced by natrojarosite, amorphous ferric sulfate (viz., g-phase), and ferric oxyhydroxide, respectively. We presume that a following transforming reaction had occurred to natrojarosite as a result of pH change before completion of the layered structure: $2NaFe_3(SO_4)_2(OH)_6 + 6Fe^{3+} + 18OH^- + nH_2O \rightarrow Na_2SO_4 + 3Fe_4(SO_4)(OH)_{10} \cdot nH_2O$. We then postulate that a precursor of the newly found structure had a three-fold structure from core to rim, $Na_2SO_4$, g-phase, and ferric oxyhydroxide. Upon a shock loading, the outermost two layers must have dehydrated into the symplectite and iron oxide, resulting in the observed layered structure[30].

A Na-sulfate[31] found in a carbon-rich clast in LAP 02342, a CR2-type meteorite, also exhibits an oxygen isotopic composition similar to the Acfer-094 symplectite. The observation of such a Na-sulfate more than once implies the sulfate ion as a carrier of heavy O-isotopes. The authors[31] suggested that Na should come from silicates in the meteorite.

*The second line of evidence*

Molecular emission spectroscopy of OCS molecule toward the Orion KL cloud[32] have revealed its $^{16}O/^{18}O = 250 \pm 135$, about twice as enriched in the heavier isotope relative to the terrestrial abundance ratio (498.70 in SMOW), which is translated to $\delta^{18}O$ ~1000 ‰. Large variations, $^{16}O/^{18}O = 25$-$330$ and $20$-$190$ for $SO_2$ and SO, respectively, have been detected[33] toward different components in the same cloud, corresponding to $\delta^{18}O$ ~ 20000-500 ‰ and ~ 25000-1600 ‰, respectively. The $^{16}O/^{17}O$ in $SO_2$ was also determined[33] for the same components, 70-670 (cf. 2632.3 in SMOW), corresponding to $\delta^{17}O$ ~ 35000-2900 ‰. Although large errors are attached to the observed values, they indicate that oxygen in these molecules in the molecular cloud is enormously enriched in $^{18}O$ and $^{17}O$ relative to $^{16}O$.

The mass spectrometric measurement of the coma of comet 67P/Churyumov-Gerasimenko has revealed oxygen isotopic compositions[34] of SO, $SO_2$, OCS, and $H_2CO$, their $^{16}O/^{18}O = 239\pm52$, $248\pm88$, $277\pm70$, and $256\pm100$, respectively, corresponding to $\delta^{18}O = 800$-$1090$ ‰. Comets are believed to have inherited the physical and chemical states of solid matter in their parent molecular cloud since they were kept at very low temperatures even in the protoplanetary disk judging from their icy volatiles such as CO, $CH_4$ and $N_2$. That these four molecules exhibit similar oxygen isotopic compositions, $^{16}O/^{18}O$ ~260 or $\delta^{18}O$ ~1000 ‰, suggests a common source for their oxygen. Atomic oxygen is the most plausible common oxygen, derived from the self-shielding of CO molecules, with its $\delta^{18}O$ ~1000 ‰. The flux of UV photons through the molecular cloud, that were able to dissociate $^{12}C^{16}O$, must have been reduced to half (260/527 = 0.49, where 527 is Sun's $^{16}O/^{18}O$ ratio), whereas the UV fluxes for dissociating $C^{18}O$ and $C^{17}O$ were not attenuated.

Recently it was found that the Fe-Ni sulfide in the Acfer-094 symplectite had mass-independent fractionation effects in its sulfur isotopes[35], that were attributed to the photochemical dissociation of $H_2S$ gas molecules in our molecular cloud by intense UV from a nearby O-, B-star association. We suggest that atomic S of the $H_2S$-dissociation origin and atomic O of the CO-dissociation origin met on icy grains in the same molecular cloud and subsequently formed SO molecules. This would explain why the Acfer-094 symplectite has its characteristic isotopic compositions both in O and S isotopes.

A relatively high temperature of dust grains (>70 K), however, was invoked[35] in the cloud so that the $H_2S$ ice evaporated completely to become gaseous $H_2S$. The temperature appears too high because not only H atoms but also O atoms would not adsorb on the $H_2O$ ice surface[36], implying that the atomic O of the CO-dissociation origin could not be quenched as $H_2O$ and also as SO molecules. Whether H and O atoms can stay long on the dust surface until they meet other atoms and molecules depend strongly on dust temperature and dust species. So we have calculated the sublimation temperature ($T_d$) of $H_2S$ ice in molecular clouds as a function of number density ($n_H$) and temperature ($T_g$) of the gas, and obtained 44.0-47.7 K for $T_d = T_g$ and $n_H = 10^4$-$10^6$ (cm$^{-3}$) (see **Materials and Methods**), significantly lower than 70 K and reassuring the SO formation because of a relatively high desorption energy for atomic O on the $H_2O$ ice, ~800 K[37].

However, H atoms have low desorption energies[37], 450 K and 650 K, on crystalline and amorphous $H_2O$ ices, respectively. The surface state, e.g., roughness, porosity, and dislocation density, may also affect adsorption/desorption kinetics[37]. Therefore, it is not certain that the temperature consistent with the S isotopic compositions, > 44-48 K, is low enough for H atoms to be kept on the icy grains for a long duration so as to meet O atoms coming from the gas phase.

For the moment, we tacitly presume that a major carrier of $^{17,18}$O-enriched oxygen is water even in Acfer 094 meteorite because water is the most abundant O-bearing species and because SO and $SO_2$ are minor species for the small abundance of S relative to O and H[38]. Due to the inherently high abundance of $H_2O$ ice in the molecular cloud, the addition of $H_2O$ ice with its oxygen of the CO-dissociation origin would not change much the oxygen isotopic composition of the bulk $H_2O$ ice. For example, if a heavy $H_2O$ ice having its $\delta^{18}O \sim \delta^{17}O \sim$ 1000 ‰ were mixed with the indigenous ice in the molecular cloud having $\delta^{18}O \sim \delta^{17}O \sim$ -60 ‰, viz. that of the sun, only 7.6 percent would be needed for the proportion of the heavy ice to the bulk ice in order to explain the isotopic composition of the bulk ice, $\delta^{18}O \sim \delta^{17}O \sim$ 20 ‰, viz. the maximum values for the initial water described above[11]. This would explain nearly terrestrial oxygen isotopic compositions for most of the solar system materials except for the Acfer-094 symplectite (Fig. 1).

It is a reasonable assumption that Acfer 094 inherited the same SO and $SO_2$ molecules with the same oxygen isotopic composition as those in comet 67P/C-G, because both solar system materials evolved from the common molecular cloud. Then, why does the Acfer-094 symplectite (magnetite) show $\delta^{18}O$ = 180 ‰ instead of 1000 ‰? The question appears easy to answer because magnetite ($Fe_3O_4$) and its precursory sulfate ion ($SO_4^{2-}$) has four oxygens. If its three oxygens other than that of the SO origin had a near-normal isotopic composition, ~ 0 ‰, the $SO_4^{2-}$ would have ~ 250 ‰ on average, not very far from 180 ‰. A source of oxygen with a near-normal composition is considered.

Acfer 094, like other meteorites, is a debris of once-large solar system body, planetesimal, which accumulated cold solid grains distributed in the protoplanetary disk. If turbulence in the disk was large enough, diffusive mixing[39] of dust grains in the vertical direction was so effective that icy ($H_2O$) grains less than a micrometer were lifted to the disk upper layer where $H_2O$ ice was photo-destructed[40] to H + OH. We postulate that the OH radicals met SO and $SO_2$ molecules of the molecular cloud origin in the $H_2O$ ice and reacted successively[41]: SO + OH → $SO_2$ + H, $SO_2$ + OH → $HSO_3$, and $HSO_3$ + OH → $H_2SO_4$.

Starting from the SO, the oxygen isotope of $H_2SO_4$ (and of Acfer 094 magnetite) must be a 1 : 3

mixture of the oxygens, one from the SO molecule ($\delta^{18}O \sim \delta^{17}O \sim 1000$ ‰) and three from the $H_2O$ ice ($\delta^{18}O \sim \delta^{17}O \sim 20$ ‰, assumed), exhibiting $\delta^{18}O \sim \delta^{17}O \sim 265$ ‰ on average (Fig. 2b). Starting from the $SO_2$ ($\delta^{18}O \sim \delta^{17}O \sim 1000$ ‰), on the other hand, the oxygen in $H_2SO_4$ must be a 1 : 2 mixture of $SO_2$ and $H_2O$, resulting in $\delta^{18}O \sim \delta^{17}O \sim 510$ ‰ (Fig. 2b).

Another route of $H_2SO_4$ formation has been suggested by the experiment[42] where the mixture of $H_2O + SO_2 + H_2O_2$ ices subjected just above ~100K for more than 100 minutes produced $H_2SO_4$ by thermally-induced reactions. The presence of $H_2O_2$ in the protoplanetary disk was very likely. In the above scenario of OH formation from $H_2O$ ice, the supply of photo-produced OH into the icy grains in the protoplanetary disk must have been more than enough to form $H_2SO_4$ from SO or $SO_2$. Unless there were other atoms or molecules on the ice, sufficient and mobile enough to combine with OH, excess OH should have been quenched in the ice, either in the lattice of or interstitially between $H_2O$ grains. Upon warming of the icy grains (>90 K), mobile OH molecules would form $H_2O_2$ ice[43,44]. Accordingly the $H_2O_2$ ice must have the same oxygen isotopic composition as that of $H_2O$ ice.

In either case, the oxygen isotopic composition of $H_2SO_4$ is significantly different from that of the Acfer-094 symplectite, 180 ‰. A major conciliation to this problem is given later.

*The third line of evidence*

Generally, aqueous alteration of matrix materials in carbonaceous meteorites is thought to have occurred above 0 °C (up to over 150 °C) for prevalent occurrences of phyllosilicates such as serpentine and for the oxygen isotopic composition of carbonates in the meteorites[11,12]. Acfer 094, on the other hand, has no phyllosilicate (less than 1% of its volume)[45] while amorphous silicates in its matrix contain fairly abundant water[46,47] (3-15 % by weight). Temperature-pH condition of the coexisting liquid is a key to this apparent inconsistency. Serpentinization of Mg-rich minerals such as olivine and pyroxene occurs at high pH (~13) at low temperatures (as low as ~0 °C), while it proceeds in weakly alkaline conditions (pH ~10) even at high temperatures (~100 °C)[48].

Sulfuric acid provides conditions of low temperature and low pH simultaneously. If the bulk composition in the system $H_2SO_4$-$H_2O$ is assumed to be 10-20 wt.% $H_2SO_4$ or 1.0-2.0 M (a predicted concentration for the fluid in the Acfer 094 parent planetesimal in the later discussion to follow), the liquidus temperature, viz. the temperature for total melting of ice, is -5 to -15 °C (Fig. 3). The composition of the ice-SAH (sulfuric acid hemihexahydrate) eutectic (-62 °C) is 36 wt.% $H_2SO_4$ or 4.6 M (Fig. 3). Accordingly, the sulfuric acid concentration in the liquid varies from 4.6 M at the eutectic (-62 °C) to 1.0-2.0 M at the liquidus (-5 to -15 °C). The pH values corresponding to these conditions are $H_o$ (Hammett acidity)[49] = -2.77 (-62 °C) and pH ~ $H_o$ = -0.41 (-5 °C) to -1.205 (-15 °C).

A highly porous amorphous silicate lithology[47] found in the Acfer 094 matrix was interpreted as a pre-existence of 100-500 nm sized icy grains. Accordingly, it is suggested that the Acfer 094 was kept below the liquidus temperature (viz. < -5 °C or < -15 °C) during its parent planetesimal era. Thus, the fluid system $H_2SO_4$-$H_2O$ prevented Mg-rich minerals from serpentinization. Instead, acids must have promoted dissolution of silicates to colloidal sol; this issue will be discussed later.

As discussed above, the ice in the protoplanetary disk likely contained not only $H_2SO_4$ but also $H_2O_2$. We postulate that in the Acfer 094's parent planetesimal metal ($Fe^0$) reacted with sulfuric acid, forming $Fe^{2+}$ by reaction: $Fe^0 + 2H^+ + H_2O_2 \rightarrow Fe^{2+} + 2H_2O$ --- [R1]. In R1, $H_2O_2$ works as a depolarizer to remove nascent hydrogen from the metal surface[50], thereby promoting the reaction continuously. $H_2O_2$ prompts a further oxidation of $Fe^{2+}$ to $Fe^{3+}$: Fenton process[51] is a very efficient system of iron redox reactions because of radical formation from $H_2O_2$, and effective in acidic conditions. $Fe^{2+} + H_2O_2 + H^+ \rightarrow Fe^{3+} + {}^*OH + H_2O$ --- [R2], where asterisk denotes radicals. Additionally, $Fe^{2+} + {}^*OH + H^+ \rightarrow Fe^{3+} + H_2O$ --- [R3]. Reaction R1 takes ~40 sec. to complete dissolution of an iron grain with radius 1 μm immersed in 1 M $H_2SO_4$ at -10°C, calculated from the experimental data[50]. Reaction R2[51] takes only ~0.01 sec if the concentration of both $Fe^{2+}$ and $H_2O_2$ is assumed 1 M. Radical reaction R3 is even faster. Consequently, as soon as $Fe^{2+}$ forms by reaction R1, it is immediately oxidized to $Fe^{3+}$.

$H_2O_2$ should have also attacked Fe-Ni sulfide in Acfer 094 (pyrrhotite $(Fe,Ni)_{1-x}S$, x=0-0.2 and pentlandite $(Fe,Ni)_9S_8$), producing an extra $SO_4^{2-}$ by the overall reactions[52]: $Fe_{0.875}S + 2\ Fe^{3+} + 2.875\ H_2O_2 \rightarrow 2.875\ Fe^{2+} + SO_4^{2-} + 2.25\ H^+ + 1.75\ H_2O$ --- [R4], where x = 0.125 for a typical pyrrhotite, $Fe_7S_8$. No Ni is assumed for brevity. Many experiments have exhibited a concurrence of elemental sulfur ($S^0$) according to: $Fe_{0.875}S + 0.875\ H_2O_2 + 1.75\ H^+ \rightarrow 0.875\ Fe^{2+} + S^0 + 1.75\ H_2O$ --- [R5], which supersedes R4 as the reactions proceed and tends to halt all reactions due to the formation of a $S^0$-rich layer[52,53]. Reaction R4 takes ~1000 sec to complete dissolution of a pyrrhotite grain with radius 1 μm at -10 °C, pH = 2.5, and $Fe^{3+} \sim H_2O_2$ = 1 M, calculated from the experimental data[52]. Therefore, iron sulfides dissolve more slowly than iron metal.

The constituents needed for formation of the g-phase, $Fe^{3+}$, $SO_4^{2-}$, $OH^-$ and $H_2O$, are readily available in the $H_2SO_4$-$H_2O_2$-$H_2O$-$Fe^0$-FeS system. From the viewpoint of oxygen isotopes, dissolution of sulfides like in reaction R4 is important as it produces $SO_4^{2-}$ ions with oxygen isotopes of the $H_2O_2$ or $H_2O$ origin, viz. $\delta^{18}O \sim \delta^{17}O \sim$ 20 ‰. This secondary $SO_4^{2-}$ formed in the planetesimal cryofluid should mix with the primary $SO_4^{2-}$ of the molecular-cloud and protoplanetary-disk origin which exhibits $\delta^{18}O \sim \delta^{17}O \sim$ 265 ‰ or 510 ‰, depending on its original source, SO or $SO_2$. Their mixing ratio should determine the final oxygen isotopic composition of $SO_4^{2-}$, hence of $Fe_3O_4$ in the Acfer-094 symplectite ($\delta^{18}O \sim \delta^{17}O \sim$ 180 ‰). In case of SO, the mixing ratio of the secondary $SO_4^{2-}$ to the primary $SO_4^{2-}$ should be ~0.5; in case of $SO_2$, ~2 (Fig. 2c).

*The fourth line of evidence*
The four oxygen atoms in $SO_4^{2-}$ are so strongly bonded to its central S atom that $H_2SO_4$ hardly exchanges its oxygen (and sulfur) isotopes with other substances. According to the isotope exchange experiment[54], the half-life of isotopic exchange for sulfate ions with water in sulfuric acid solution with pH = 2.8 (or 5.0) is calculated as $2 \times 10^5$ (or $10^9$) yrs at 0 °C and $10^{10}$ (or $10^{14}$) yrs at -60 °C. It means that the oxygen, once locked in sulfate ions, keeps its isotopic composition unless they are subjected to high temperatures (>100 °C) for long duration (>$10^6$ yrs) at pH >5.

Thus, sulfate ions in the g-phase precipitated from ferric sulfate solutions should have the same oxygen isotopic composition as that of sulfuric acid. As already described, magnetite and sulfide in the Acfer-094 symplectite which formed by dehydration and decomposition of the g-phase should inherit the same isotopic compositions (both of oxygen and sulfur) as those of the sulfate ions. Magnetite is also known for its slow diffusion rate of oxygen under hydrothermal conditions[55].

Therefore, its isotopic composition must have been kept intact after shock-fragmentation of the Acfer 094's parent planetesimal, and also survived atmospheric entry and a possible weathering in the Sahara Desert, that the Acfer 094 meteorite experienced.

Ironically, the three oxygen atoms in sulfite ion, $SO_3^{2-}$ or bisulfite ion, $HSO_3^-$, are so weakly bound to the S atom that they readily exchange with water oxygen[56]. Following the sulfite-water isotope exchange experiment[56], the half-life of isotopic exchange for sulfite ions with water in the solution with pH = 2.8 (or 5.0) is calculated as $5 \times 10^{-10}$ (or $10^{-5}$) sec at 0 °C and $5 \times 10^{-8}$ (or $10^{-3}$) sec at -60 °C. Therefore, the isotopic composition of both O (and S) of sulfite ions would be immediately set equal to that of abundant bulk water.

The comet 81P/Wild 2 sample contained another symplectically intergrown sulfide/Fe-oxide (pentlandite/maghemite)[57]. Its oxygen isotopic composition, however, was found indistinguishable[57] from the terrestrial value, $\delta^{18}O \sim \delta^{17}O \sim 0$ ‰. We predict that the precursor of the 81P symplectite was a ferric sulfite hydroxylate with the chemical formula[58], $Fe_2(SO_3)_3 \cdot 7Fe(OH)_3$. It must have been dehydrated and symplectically decomposed by shock heating, such as: $Fe_2(SO_3)_3 \cdot 7Fe(OH)_3 \rightarrow [Fe_9(SO_3)_3]^* + 21OH$ ; $[Fe_9(SO_3)_3]^* \rightarrow 3FeS + 3Fe_2O_3$, where asterisk* denotes a transition state.

**Petrographic interrogation**

The symplectite is contained in the matrix of Acfer 094. Its modal abundance is 50-2500 ppm from location to location, with an average of only 600 ppm[13]. Its another puzzling characteristic is that it is always accompanied by a fractured sulfide grain as explained in the first line of evidence. Below we interrogate the Acfer 094 matrix to find an answer to the question why the symplectite is so small in abundance and is accompanied by a fractured sulfide. We also try to locate "missing" sulfur and iron in the matrix, which are included neither in metal, sulfides, and silicates nor in the symplectite.

*Amorphous silicates*

Amorphous silicates occupy nearly ~40 vol.% of the Acfer 094 meteorite matrix[45]. They typically occur as sub-μm rounded nodules, but often extend together with organic materials, enclosing rounded small crystals of silicates, sulfides, and minor metal, some 100-300 nm in size[45].

Here, we focus on the chemical composition of the amorphous silicates, which is uniform but typically enriched in Si, Al, and $Fe^{3+}$ and depleted in Mg, $Fe^{2+}$, Ca, Na, and S, relative to their solar elemental abundances normalized to the Si abundance[46,59]. This trend of enrichment and depletion is analogous to that of silica-rich gel precipitated from a mixed sol of dissolved silica and various ions, where multi-valent cations (e.g., $Al^{3+}$, $Fe^{3+}$, and $Cr^{3+}$) are selectively adsorbed[60,61] to silanol group of the hydrated surface of silica gel ($M^{n+} + m(-SiOH) \rightleftharpoons M(OSi-)_m^{n-m} + mH^+$, n: cation valence, and m: number of silanols) leaving behind less-charged cations ($Fe^{2+}$, $Ca^{2+}$, $Mg^{2+}$, and $Na^+$) in the liquid phase.

The rounded small crystals of silicates and sulfides enclosed by the amorphous silicate in Acfer 094 are suggestive of their corrosion in a liquid medium. Fe metal is absent in the amorphous silicate nodules[46,59], implying its complete dissolution. As we indicated in the third line of evidence, we postulate that the amorphous silicates in Acfer 094 are low-temperature colloidal precipitates from the sol in a sulfuric-acid-based cryofluid, that was previously a mixture of silicate crystals (mostly olivine and pyroxene), sulfides and metal before their corrosive dissolution. The amorphous silicates in Acfer 094 actually contain a significant amount of H$_2$O, 3-18 wt.%[47] or OH, ~10 wt.%[46]. The H$_2$O-rich fluid containing mostly unadsorbed cations (Fe$^{2+}$, Mg$^{2+}$, Ca$^{2+}$, and Na$^+$) and their charge-balancing anions (SO$_4^{2-}$ and probably Cl$^-$, NO$_3^-$, NO$_2^-$, PO$_4^{3-}$, and so on) was likely squeezed out of the gel as gelation proceeded[62].

So Na$^+$ was available just as predicted[31]. Due to a preferential adsorption of Fe$^{3+}$ onto silica-gel, however, the fluid would be depleted in Fe$^{3+}$. The situation is not favourable for precipitation of the g-phase. Worse because the abundant Mg$^{2+}$ in the fluid tends to interrupt[22] Fe$^{3+}$ and SO$_4^{2-}$ ions from making a complex such as FeSO$_4^+$ and Fe(SO$_4$)$_2^-$, a prelude to a successful precipitation of g-phase. This may be the circumstance that the g-phase, the precursor of the symplectite, could not easily precipitate from the cryofluid.

Nonetheless, the dissolution of silicates must have played an important role in formation of the g-phase because its precipitation requires[14,19,22] OH$^-$ ions, or pH to increase more than 3. When silicates dissolve in an acidic solution, the hydrogen ion is continuously consumed: Mg$_2$SiO$_4$ + 4 H$^+$ → Si(OH)$_4$ + 2Mg$^{2+}$ and MgSiO$_3$ + 2H$^+$ + H$_2$O → Si(OH)$_4$ + Mg$^{2+}$. This would shift pH toward higher values. While at the start the corrosion reactions of metal and sulfides dominate over the silicate dissolution, they should soon decline because the Fenton process[51] becomes less effective at pH above 3.5 and all corrosion reactions (R1-R5) exhaust a key chemical, H$_2$O$_2$. Thus, with corrosion of metal and sulfides declining, the dissolution of silicates gradually takes control of the pH change. It must be this time when the g-phase started to precipitate from the fluid.

On the other hand, keeping Fe$^{3+}$ out of silanol's reach is probably another key. We predict that it is provided by a metal-sulfide composite (before its evolving to the pair of symplectite and fractured sulfide by shock) because above the pH 2.7, the isoelectric point of pyrrhotite, the Fe$^{3+}$ produced in Fenton process from a coexisting metal is trapped by the negatively-charged pyrrhotite surface, while transformation of Fe$^{3+}$ into Fe$^{2+}$ by reaction R4 becomes suddenly ineffective[52] above this pH. Solid elemental sulfur produced by reaction R5 is also negatively charged, thus providing nuclei for a Fe$^{3+}$-bearing phase to precipitate, perhaps for natrojarosite first, then g-phase, explaining the elemental sulfur that coexists with Na$_2$SO$_4$ in the layered structure of the newly found Acfer-094 symplectite[30]. The low abundance of the symplectite may be simply ascribed to a paucity of the metal-sulphide composite.

*Fe and S budgets*

According to their modal abundance, the Acfer 094 matrix is very poor in metal and sulfides, only 2.3 vol.% combined[63]. The volume % of metal and sulfides (normal and fractured sulfides combined) in the Acfer 094 matrix can be separately estimated from their cumulative size distributions[13], ~0.5 vol.% and ~2 vol.%, respectively. Assuming the average density of the matrix ~2.8 g/cm$^3$ and those of the phases (7.874 and 4.6 g/cm$^3$, respectively), the elemental abundances of Fe and S contained in these phases account for only 3.3 wt.% and 1.1 wt.%, respectively. These numbers are very small compared to those, 24.0 wt.% Fe and 7.2 wt.% S, in water-free Orgueil meteorite[64], a reference material in the solar system. Whereabouts of the rest of the elements are considered assuming that the Acfer 094 matrix contained the same amounts of Fe and S as a water-free Orgueil.

The rest of sulfur other than sulfides, viz. ~85 % of its total, must exist as $SO_4^{2-}$ and elemental sulfur from the preceding discussion. If dissolution of sulfides (R4 and R5) generated a 1 : 2 mixture of the secondary $SO_4^{2-}$ and elemental sulfur, and if the mixing ratio of the secondary to the primary $SO_4^{2-}$ was 2, implying $SO_2$ as the original oxygen isotope carrier, the total $SO_4^{2-}$ would account for 36 % of the total sulfur. If, on the other hand, sulfides dissolved into a 2 : 1 mixture of $SO_4^{2-}$ and elemental sulfur, the total $SO_4^{2-}$ would be 64 % of the total sulfur for the same secondary/primary $SO_4^{2-}$ ratio, viz., 2. A simple calculation of the molar ratio of $SO_4^{2-}$ to $H_2O$ from these two numbers using the solar abundance[38] for $H_2O$ gives the concentration of $H_2SO_4$, 10-20 wt.% or 1-2 M, in the cryofluid of Acfer 094's parent planetesimal. A similar calculation can be done for SO as the original isotope carrier.

The rest of iron other than metal and sulfides, viz. ~86 % of its total, should include $Fe^{2+}$ and $Fe^{3+}$ ions in the cryofluid, in addition to the iron in silicates in the matrix. The silicates in the Acfer 094 matrix consist of amorphous silicates, ~40 vol.% of the matrix, and crystalline silicates, ~50 vol.%[45]. We consider that most of the iron in the amorphous silicate were once-dissolved ions from metal and sulfides, which were subsequently trapped by silanols during gelation to the amorphous silicate; they account for 34 % of the total iron as calculated from the iron content in the amorphous silicate[46]. The crystalline silicates account for only ~1 % of the total iron because olivine and pyroxene[45] are very poor in Fe. The contribution by the symplectite ($FeS/Fe_3O_4$) is negligible. By subtraction, ~50 % of the total iron must have existed in the form of $Fe^{2+}$ and $Fe^{3+}$ ions together with charge-balancing anions (e.g., $SO_4^{2-}$) in the cryofluid.

As explained before, the gelation of colloidal silicate squeezes out the interstitial fluid containing unadsorbed cations ($Fe^{2+}$, $Mg^{2+}$, $Ca^{2+}$, and $Na^+$) and their charge-balancing anions ($SO_4^{2-}$, $Cl^-$, $NO_3^-$, etc.). Therefore, the cryofluid must have been rich both in cations and anions. We presume that during the impact evaporation of volatiles (mostly $H_2O$) from the Acfer 094 planetesimal, refractory cations and anions precipitated as extremely fine-grained salt particles. These salts, notably sulfates and possibly nitrates, would have kept their oxygen isotopic compositions similar to that of the Acfer-094 symplectite, but are highly soluble in water, rendering the Acfer 094 meteorite very susceptible to weathering in terrestrial environments and also in laboratories.

**Conclusion**

The oxygen isotope in the Acfer-094 symplectite traces back its origin to the SO or $SO_2$ molecules that evolved in our molecular cloud. Following their transformation into a sulfuric acid in the protoplanetary disk, the $H_2SO_4$-$H_2O_2$-$H_2O$ fluid furnished an acidic and oxidizing cryofluid environment in the Acfer 094 parent planetesimal. An extensive corrosion of metal, sulfides, and silicates gave rise to a colloidal mixture of various cations, anions, complexes, and polymers, from which an amorphous ferric hydroxysulfate hydrate with the chemical formula $Fe_4SO_4(OH)_{10}\cdot nH_2O$ (n=1-3) managed to precipitate before the shock heating. Owing to a rigid structure of $SO_4^{2-}$ ion, its isotopic signatures were safely transferred to the magnetite and sulfide in the symplectite. It is concluded that the Acfer-094 symplectite and its progenitor is strongly coupled with the material evolution in the solar system since the days of our molecular cloud.

**Acknowledgements** We thank K. Kuramoto and the members of the Planetary and Space Science group, Hokkaido University for their continuous and sincere support.

**Author contributions** A.H. and Y.N. conceived the basic concept. A.H. developed supporting theoretical models and Y.N. surveyed literature data. A.H. wrote the manuscript with Y.N.'s help. Both authors contributed to the final conclusion.

**Competeing interests** The authors declare no competing interests.

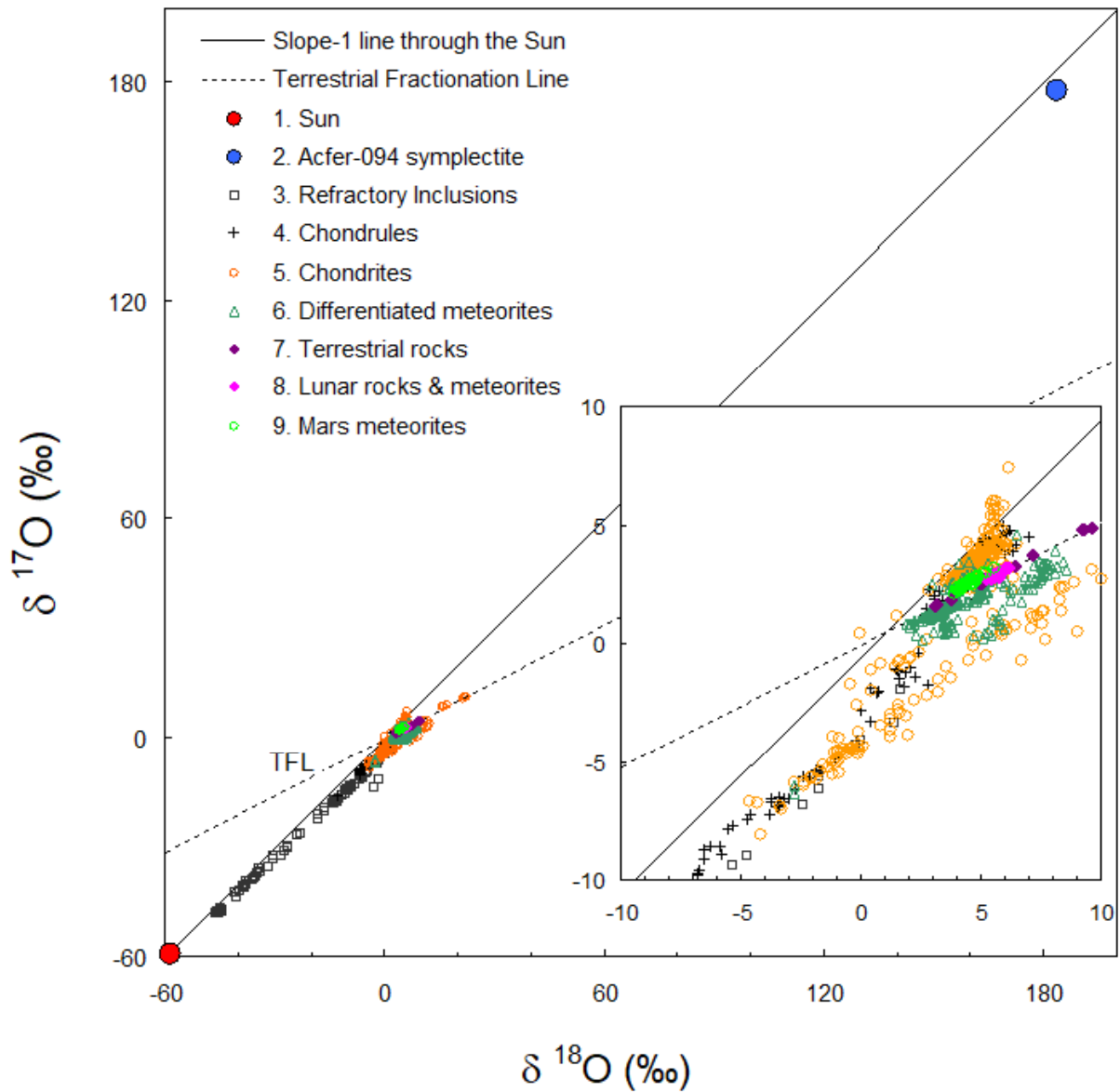

**Fig. 1** Oxygen isotopic compositions of the Sun (notation 1 in the figure)[1], planetary and preplanetary rocks, and components thereof (notations 2 through 9)[6,65,66]. Data for 2-9 are based on bulk isotope analyses with various methods. Data for individual mineral analyses are not shown for the sake of clarity. A slope-1 line through the solar isotopic composition and the terrestrial mass-fractionation line (TFL, slope = 0.52 through the origin, i.e., SMOW) are also shown for guidance.

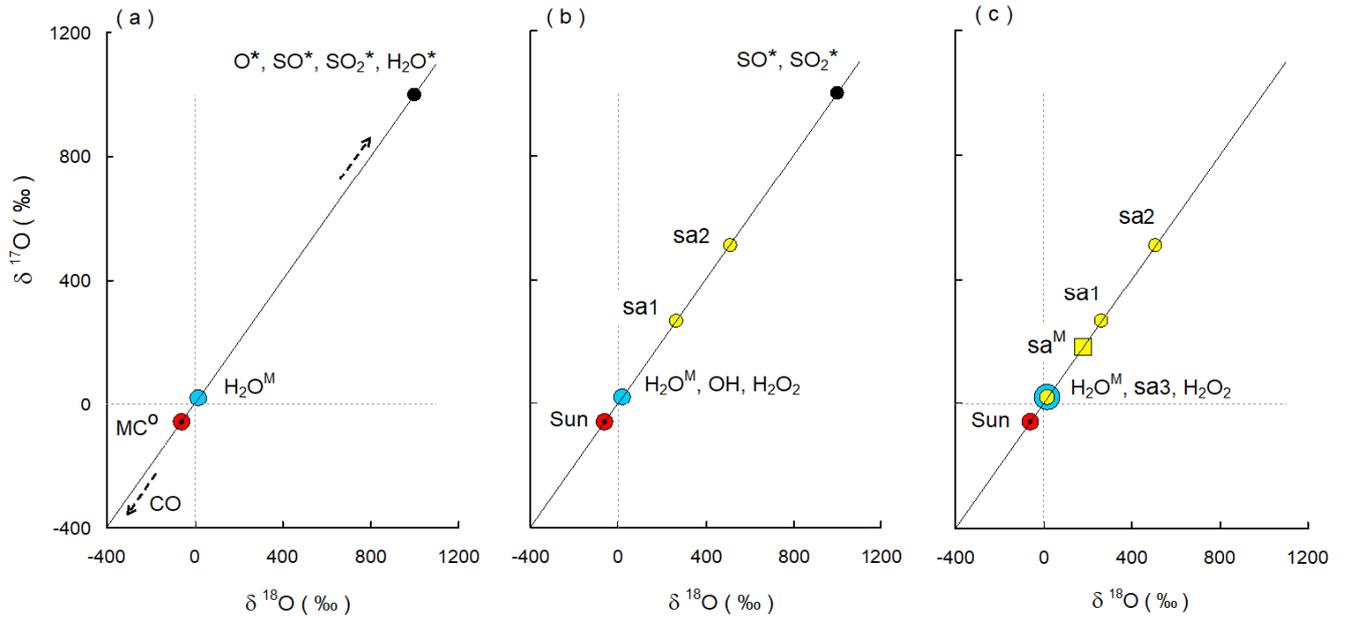

**Fig. 2** Model of evolution of the oxygen isotopic composition of sulfuric acid in the solar system. (a) Parent molecular cloud stage. Due to a self-shielding of CO molecules against external UV sources, the cloud with an originally uniform oxygen isotopic composition, MC° diversified into the $^{16}$O-rich CO gas molecules, the $^{17,18}$O-rich O atoms (denoted by asterisk; $\delta^{18}O = \delta^{17}O = 1000$ ‰), and the rest either in the gas or solid state (mostly H$_2$O ice and silicates) that kept the original composition MC°. The atomic oxygen reacted with S and H atoms on the surface of icy grains and became SO* or SO$_2$* and H$_2$O*, respectively, having the same isotopic composition. The H$_2$O* molecules were incorporated by the basal H$_2$O ice (MC°) into a mixed ice with composition H$_2$O$^M$. (b) Protoplanetary disk stage. Photo-decomposition of H$_2$O$^M$ ice grains in a disk upper layer generated abundant OH and H$_2$O$_2$ molecules with composition H$_2$O$^M$, which sequentially reacted with SO* and SO$_2$* to form sulfuric acid, H$_2$SO$_4$, having the compositions sa1 from SO* and sa2 from SO$_2$*. The protosun dictates MC° in its isotopic composition. (c) Parent planetesimal stage. In a cryofluid environment provided by sulfuric acid under the planetesimal surface, iron sulfides and ferric irons reacted with H$_2$O$_2$, forming a sulfuric acid sa3 with composition H$_2$O$^M$. The sulfuric acids of three different origins, sa1, sa2, and sa3, mixed in an appropriate proportion, gave rise to the sulfuric acid sa$^M$ with $\delta^{18}O = \delta^{17}O = 180$ ‰, which the Acfer-094 symplectite should have inherited.

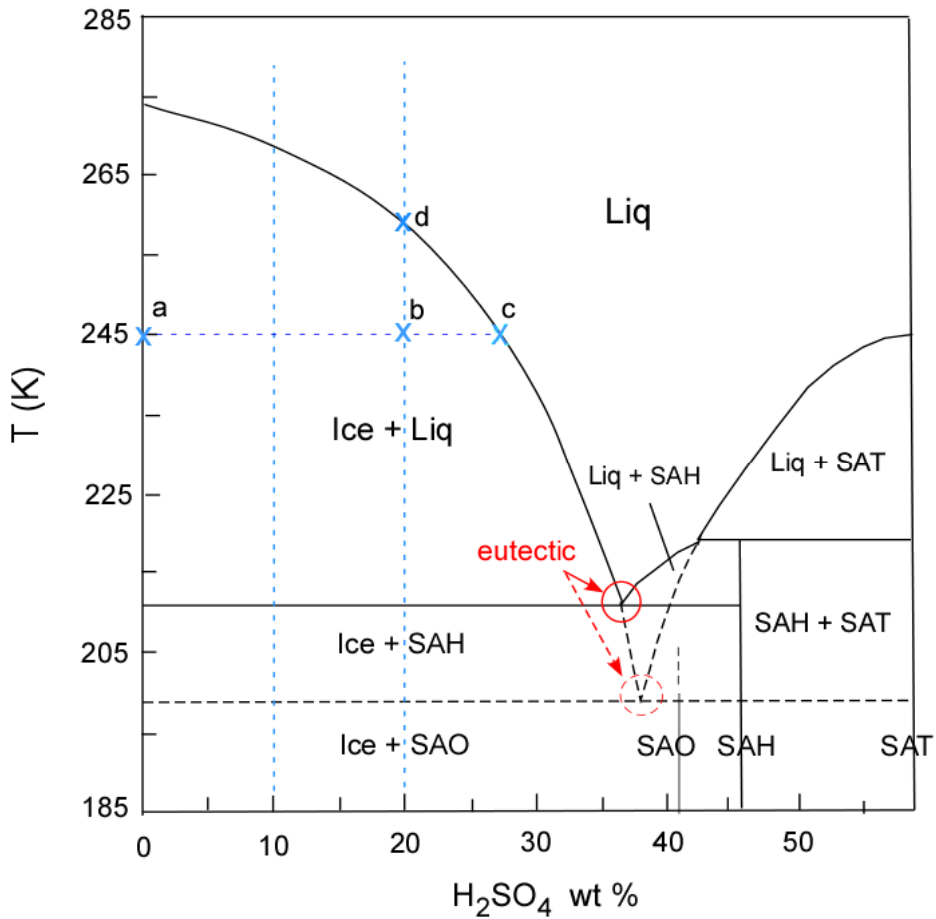

**Fig. 3** Phase relations in the $H_2O$-$H_2SO_4$ system. The composition range $H_2O$-SAT, modified after the work[67], is shown. Liq: $H_2O$-$H_2SO_4$ liquid mixture. Ice: $H_2O$ ice. SAT: sulfuric acid tetrahydrate. SAH: sulfuric acid hemihexahydrate. SAO: sulfuric acid octahydrate. Two eutectics, ice-SAH and ice-SAT, are (211 K, 36 wt.% $H_2SO_4$) and (199 K, 38 wt.% $H_2SO_4$), respectively. The range, 10-20 wt.% $H_2SO_4$, is postulated to be the composition of the cryofluid that were at work in the Acfer 094's parent planetesimal (see text). Points a, b, and c denote the phase relation at 245 K and 20 wt.% for example. Point c: the composition of the liquid. A line segment ratio, ab/bc, is equal to the mass ratio of the liquid to ice. Point d: liquidus temperature, 258 K or -15°C, at 20 wt.% $H_2SO_4$.

**Materials and Methods**

*Sublimation temperature of $H_2S$ ice*

We calculate the sublimation temperature of $H_2S$ ice in the parent molecular cloud of our solar system. The vapor pressure of $H_2S$ ice is given by the empirical equation[68] for low temperatures, log $P_{H2S,ice}$/(dyn/cm$^2$) = 11.637 - 1175.6/(T/K). We assume the element abundance in the molecular cloud, A(element), to be equal to the solar abundance[38] of elements, which gives log A(S)/A(H) = -4.812 where S and H are sulfur and hydrogen. Assuming that all the sulfur exist as $H_2S$, this value is identical to the number ratio of $H_2S$ molecules to $n_H$ (number/cm$^3$), the number density of hydrogen in the molecular cloud. A complete evaporation of $H_2S$ ice ensues above its sublimation temperature $T_d$, giving its pressure in the cloud, $P_{H2S,MC}$ (dyn/cm$^2$) = $n_H k_B T_g$ A(S)/A(H), where $T_g$: gas temperature and $k_B$: Boltzmann constant. Unless $T_d = T_g$, $P_{H2S,ice}$ and $P_{H2S,MC}$ are not equal. Therefore, the rate equations must be applied. The sublimation rate of $H_2S$ ice is given by $J_v$ (number/cm$^2$sec) = $(2\pi m_{H2S} k_B T_d)^{-0.5}$ $P_{H2S,ice}$ -- eq1, where $m_{H2S}$ is the mass of $H_2S$ molecule, while the collision rate to the ice is given by $J_c$ (number/cm$^2$sec) = $\alpha_{ac}$ $(2\pi m_{H2S} k_B T_g)^{-0.5}$ $P_{H2S,MC}$ -- eq2, where $\alpha_{ac}$ is accommodation coefficient. By equating eq1 to eq2, $P_{H2S,ice} T_d^{-0.5} = \alpha_{ac} P_{H2S,MC} T_g^{-0.5}$. In a special case where $T_d = T_g$, it reduces to $P_{H2S,ice} = P_{H2S,MC}$, where we assumed $\alpha_{ac} = 1$.

For $n_H$ = 10$^4$, 10$^5$, and 10$^6$ (number/cm$^3$) in the cloud, $T_d = T_g$ = 44.1, 45.8, and 47.7 K, respectively. Even when $T_g$ is higher than $T_d$ (*e.g.*, $T_g$ = 200 K), $T_d$ = 44.7, 46.4, and 48.4 K, respectively, for the same $n_H$ values. Therefore, the sublimation temperature of $H_2S$ ice hardly exceeds 50 K unless for much larger $n_H$ or $T_g$. If other S-bearing species exist, it would lower the $H_2S$ abundance, rendering the $H_2S$ sublimation temperature lower than the above values. The same is true if $\alpha_{ac}$ <1.